\documentclass[aps,prb,twocolumn,superscriptaddress,longbibliography,10pt]{revtex4-2}

\usepackage{graphicx,bm,amssymb,amsmath,xcolor,pifont,soul,multirow,array}
\usepackage[linktocpage=true,colorlinks=true,pdfborder={0 0 0},linkcolor=blue,citecolor=blue,filecolor=yellow,urlcolor=blue,bookmarks,pdfauthor={},]{hyperref}
\usepackage[normalem]{ulem}
\usepackage{epstopdf}
\usepackage{epsfig}
\usepackage{blkarray}
\usepackage{chemformula}

\def\bk{{\bf k}}

\def\bq{{\bf q}}

\def\>{\rangle}
\def\<{\langle}
\usepackage{amsmath}

\setstcolor{red}

\newcommand{\CT}[1]{\textcolor{black}{#1}}
\newcommand{\AK}[1]{\textcolor{black}{#1}}

\definecolor{dgreen}{rgb}{0.0, 0.8, 0.6}

\newcommand{\DG}[1]{\textcolor{black}{#1}}

\usepackage{titlesec}
\usepackage{lipsum}

\titleformat{\subsubsection}
  {\normalfont}{\thesubsubsection}{1em.}{}
\renewcommand{\thesubsubsection}{\roman{subsubsection}}

\begin{document}
\title{High-$T_{\rm c}$ Ag$_x$BC and Cu$_x$BC superconductors accessible via topochemical reactions}

\author{Daviti Gochitashvili}
\affiliation{Department of Physics, Applied Physics, and Astronomy, Binghamton University-SUNY, Binghamton, New York 13902, USA}
\author{Charlsey R. Tomassetti}
\affiliation{Department of Physics, Applied Physics, and Astronomy, Binghamton University-SUNY, Binghamton, New York 13902, USA}
\author{Elena R. Margine}
\affiliation{Department of Physics, Applied Physics, and Astronomy, Binghamton University-SUNY, Binghamton, New York 13902, USA}
\author{Aleksey N. Kolmogorov}
\email{kolmogorov@binghamton.edu}
\affiliation{Department of Physics, Applied Physics, and Astronomy, Binghamton University-SUNY, Binghamton, New York 13902, USA}
\date{\today}

\begin{abstract}
Hole-doping of covalent materials has long served as a blueprint for designing conventional high-$T_{\rm c}$ superconductors, but thermodynamic constraints severely limit the space of realizable compounds. Our {\it ab initio} results indicate that metastable Ag$_x$BC and Cu$_x$BC phases can be accessed via standard topochemical ion exchange reactions starting from Li$_x$BC precursors. Unlike all known stoichiometric layered metal borocarbides, the predicted AgBC and CuBC derivatives, comprising honeycomb layers bridged by dumbbells, are metallic rather than semiconducting. Anisotropic Migdal-Eliashberg analysis reveals that the intrinsically hole-doped AgBC possesses a unique combination of electronic and vibrational features to exhibit two-gap superconductivity above 50 K.

\end{abstract}	

\maketitle

\section{Introduction}
\label{sec:introduction}

The search for new high-$T_{\rm c}$ superconductors has yielded a large number of theoretical predictions but few experimental validations~\cite{Flores-Livas2020,Boeri2022}. This disconnect reflects an inverse correlation between thermodynamic stability and conventional superconductivity, now established across different material classes~\cite{Testardi1975,ak14,ak24,Cerqueira2024,Wang2025}. With the vast majority of proposed high-$T_{\rm c}$ superconductors being, at best, metastable at ambient conditions, high-pressure synthesis with possible quenching has enabled key discoveries~\cite{Drozdov2015,Drozdov2019} but remains impractical for scalable applications. Therefore, predictive efforts must address synthesis pathways alongside superconductor design, and soft chemistry methods offer a promising route to bypass thermodynamic constraints.

Topochemical reactions, first reported in 1931~\cite{Hertel1931, Cohen1964}, provide access to (meta)stable crystalline materials without the need for extreme temperature or pressure synthesis conditions. The selective replacement or removal of species at specific atomic sites while leaving the structural backbone essentially unchanged represents a kinetically controlled pathway for compositional modifications. Over the past decades, the approach has been used to produce a variety of new layered compounds, including metal oxides~\cite{Hosogi2008, Gwon2015, Roudebush2016, Abramchuk2017, Sethi2022}, (oxy)chalcogenides~\cite{Boller1992, Rutt2006, Martinolich2017, Sasaki2023}, and borides~\cite{Bhaskar2021, Baumler2023, Mou2025}. The resulting materials have exhibited enhanced photocatalytic activity~\cite{Hosogi2008}, potential for spin-liquid behavior~\cite{Bahrami2019}, improved electrochemical performance ~\cite{Mou2025}, and unconventional superconductivity~\cite{Li2019}.

The extensive experimental work has identified several factors governing topochemical transformations, including ion size mismatch, reactant selection, and reaction environment. Topotactic ion exchange preserves the original crystal structure and typically favors the replacement of larger cations with smaller ones due to diffusion kinetics and lattice constraints, as seen in Cu substituting Na in Na$_2$IrO$_3$~\cite{Roudebush2016,Abramchuk2017}, but the reverse has also been observed, such as Ag substituting Li in $\alpha$-Li$_2$IrO$_3$~\cite{Bahrami2019, Bette2019}. Some reactions involve rich chemistry with multiple stages~\cite{Alameda2019, Bhaskar2021,Sasaki2023}, while targeted ion exchange is often facilitated by the introduction of cations in the form of metal salts~\cite{Rutt2006, Hosogi2008}. {\it Ab initio} work has focused primarily on interpreting topochemical pathways or modeling the observed materials’ properties~\cite{Gwon2015,Si2020,Handy2022}. Moving toward {\it ab initio} materials design, Suzuki et al.~\cite{Suzuki2024} recently introduced a computational framework for identifying feasible topotactic reactions and predicted ternary wurtzite-type oxides by calculating free energies with density functional theory (DFT). Their experimental work validated the DFT-derived trends, leading to previously unreported Li-to-Ag exchange and the synthesis of multiple metastable compounds and solid solutions with controlled compositions.

The present study focuses on predicting new layered metal borocarbides, a diverse materials class unexplored with topochemical ion exchange. Over the years, LiBC~\cite{LiBC} has attracted wide attention due to early theoretical predictions suggesting that its hole-doped derivatives could exhibit high-temperature MgB$_2$-type superconductivity resulting from the coupling of the in-plane BC-$p_{x,y}$ orbitals and the bond-stretching modes of the electron-deficient BC framework~\cite{Rosner2002, Dewhurst2003}. However, despite extensive experimental efforts, superconductivity has not been observed in delithiated LiBC~\cite{Bharathi2002, Zhao2003,Fogg2003a,Fogg2003b,Fogg2006, Kalkan2019}, likely due \DG{deviations from the ideal BC morphology} arising from the removal of Li~\cite{Tomassetti2024a}. \DG{Namely, our ab initio calculations showed that layer buckling and interlayer bridging reduce the Tc by factors of $\sim$2 and $\sim$20, respectively, while the intralayer B-C bond rotation suppresses superconductivity completely. Our following} DFT analysis indicated the possibility of modulating the structure through soft chemical techniques, such as thermodynamically favorable reintercalation of Li$_x$BC with alkaline or alkaline earth metals~\cite{Tomassetti2024a}, obtaining mixed-metal Li$_x$M$_y$BC materials with signature electronic and vibrational features requisite for high-$T_{\rm c}$ superconductivity. Beyond LiBC, only two other metal borocarbides with honeycomb layers have been reported, MgB$_2$C$_2$~\cite{Worle1994}, and BeB$_2$C$_2$~\cite{Hofmann2008}, with another, NaBC~\cite{Tomassetti2024a}, previously predicted as a low-temperature ground state. Our {\it ab initio} phase stability and electron-phonon (e-ph) coupling calculations revealed that thermal deintercalation of MgB$_2$C$_2$ and NaBC under accessible experimental conditions could turn the semiconducting materials into hole-doped superconductors with $T_{\rm c}$ values in the 43-84~K range~\cite{Tomassetti2024b}. These proposed deintercalation or reintercalation pathways rely on the robustness of the BC layered framework, as LiBC has been shown to retain its honeycomb morphology at the ideal 1:1 stoichiometry at high temperatures (up to 1770~K) and low metal concentrations (down to at least $x = 0.5$)~\cite{Fogg2006, Kharabadze2023}. 

Here, we explore the kinetically protected subspace of borocarbides with honeycomb layers through topochemical reactions facilitated by the introduction of metal salts. Our DFT results show that the substitution of alkali or alkaline earth metals in known borocarbides with Cu, Ag, or Zn releases $\sim 100$ kJ/mol, comparable to the calculated reaction energy in successful topochemical syntheses of delafossites ~\cite{Abramchuk2017,Suzuki2024}. We carried out extensive structure searches to identify low-energy layered Li-Cu-BC and Li-Ag-BC derivatives and analyzed the superconducting properties of AgBC with the anisotropic Migdal-Eliashberg (aME) formalism \cite{Margine2013,Lucrezi2024}. Fermi surfaces from the signature hole-doped BC-$p_{x,y}$ states, electron-doped Ag-$s$, and BC-$p_z$ states in the stoichiometric AgBC are found to boost the $T_{\rm c}$ above 50~K. The findings suggest a promising route to a new class of ternary and, possibly, quaternary metal borocarbide superconductors with high $T_{\rm c}$.

\section{Methods}
\label{sec:methods}

To analyze stability of  M-B-C and Li-M-BC phases we employed {\small VASP}~\cite{Kresse1996}  using projector augmented wave potentials~\cite{Blochl1994} and a 500~eV plane-wave cutoff. 
As the subject of the study was layered materials, where dispersive interactions are important ~\cite{ak06,Lebegue2010,ak30,Ning2022}, we relied on the optB86b-vdW functional~\cite{optB86b}. Additionally, we reexamined the stability of the final set of predicted phases using r$^2$SCAN+rVV10~\cite{r2scan, Ning2022}, and employed HSE06~\cite{HSE06} to obtain more reliable band gap predictions for selected phases. All structures were evaluated with dense ($\Delta k \sim 2 \pi \times 0.025$~\AA$^{-1}$)  Monkhorst–Pack $\bk$-meshes~\cite{Monkhorst1976}. Thermodynamic stability values are reported primarily in eV per atom of the product, such as AgBC with three atoms per formula unit, while reaction energies are also given in kJ per mole to facilitate the comparison with previous studies (see Supplementary Note I).

Global structure optimizations were performed using an evolutionary algorithm implemented in the MAISE package~\cite{maise}. In fixed-composition runs, randomly initialized 16-member populations with up to 22 atoms per unit cell evolved for up to 250 generations through standard mutation and crossover operations~\cite{maise}. For the systematic screening of possible interlayer site decorations in metal borocarbides, we constructed various CuBC and AgBC cells and sequentially substituted and/or removed metals while retaining only the nonequivalent configurations. Equivalence was determined using our structural fingerprint based on the radial distribution function~\cite{ak16,maise}. The generated supercells included standard expansions of the fully occupied hP6 structure, recognized as the ground state for LiBC~\cite{LiBC}. These expansions comprised $\sqrt{3}\times\sqrt{3}\times1$, 2$\times$2$\times$1, and larger orthorhombic supercells with up to 54 atoms. We considered both AA and AA$^\prime$ stackings of the BC layers. In total, we examined a diverse set of layered phases, with more than 4000 unique metal arrangements for each system.

For selected structures, the thermodynamic corrections due to vibrational entropy were evaluated using the finite displacement method implemented in PHONOPY~\cite{Togo2015}. We used supercells ranging from 69 to 264 atoms, with an average of 138 atoms across 80 phases, and applied 0.1~\AA\ displacements within the harmonic approximation. \DG{ {\it Ab initio} molecular dynamics (AIMD) simulations were performed using the VASP code, with a $400$ eV energy plane-wave cutoff and $4\times4\times2$ $\bk$-mesh for the $3\times3\times2$ supercell, to evaluate the thermal stability of the predicted AgBC phase. The simulations were carried out in the {\it NVT} ensemble using a Nos\'e-Hoover thermostat at a constant temperature of $600$ K for $10,000$ $1$-fs timesteps.}

The \textsc{Quantum} ESPRESSO package~\cite{QE} was employed for calculating ground-state properties associated with superconductivity, using the optB86b-vdW~\cite{optB86b} functional and norm-conserving pseudopotentials from the Pseudo Dojo library~\cite{Dojo2018}. The lattice parameters and atomic positions were relaxed until the total energy was converged within $10^{-6}$ Ry and the maximum force on each atom was less than $10^{-4}$ Ry/\AA. A plane-wave cutoff value of 100~Ry, a Methfessel-Paxton smearing~\cite{Methfessel1989} value of 0.02~Ry, and  $\Gamma$-centered Monkhorst-Pack~\cite{Monkhorst1976} \textbf{k}-meshes were used to describe the electronic structure. The dynamical matrices, the linear variation of the self-consistent potential, and electron-phonon coupling strength were calculated within density-functional perturbation theory~\cite{Baroni2001} on irreducible sets of regular \textbf{q}-meshes given in Supplementary Table~3, \DG{shown in previous studies to ensure convergence for related layered materials~\cite{Tomassetti2024a,Tomassetti2024b, Kafle2022}.} For example, a 30$\times$30$\times$18 \textbf{k}- and 6$\times$6$\times$4 \textbf{q}-mesh were used for AgBC.

We utilized the EPW code~\cite{Giustino2007, EPW2016, Margine2013, EPW2023} to evaluate the e-ph interaction and superconducting properties of AgBC. The Wannier interpolation~\cite{WANN1, WANN2, WANN3} was performed on a uniform 6$\times$6$\times$4 $\Gamma$-centered \textbf{k}-grid with the Wannier90 code~\cite{WANN1, WANN2, WANN3} in library mode. As projections for the maximally localized Wannier functions, we used $s$ and $p$ orbitals for every C atom, and $d$ orbitals for every Ag atom to reproduce the electronic structure, totaling 9 Wannier functions. The anisotropic full-bandwidth ME equations~\cite{EPW2023,Lucrezi2024} were solved with a sparse intermediate representation of the Matsubara frequencies~\cite{Mori2024} on fine uniform 100$\times$100$\times$70 $\bk$- and 50$\times$50$\times$35 $\bq$-point grids, with an energy window of $\pm 0.4$~eV around the Fermi level. The Coulomb $\mu^*$ parameter was chosen to be 0.20 in order to ensure agreement between measured and computed aME $T_{\rm c}$ values for MgB$_2$~\cite{Kafle2022}. The EPW code was also used to find the isotropic Migdal-Eliashberg critical temperatures using the Eliashberg spectral functions outputted by the PHonon code of \textsc{Quantum} ESPRESSO. 

Visual representations of crystal structures and Fermi surfaces were respectively created with VESTA~\cite{vesta} and FermiSurfer~\cite{fermisurfer}. To resolve different phases at the same composition, we use Pearson symbols and space groups when needed. Full structural information for relevant DFT-optimized M$_y$BC or Li$_{x-y}$M$_y$BC (M = Cu or Ag) phases is provided as CIF files in the Supplementary Material.
\begin{figure}[t!]
   \centering
\includegraphics[width=0.48\textwidth]{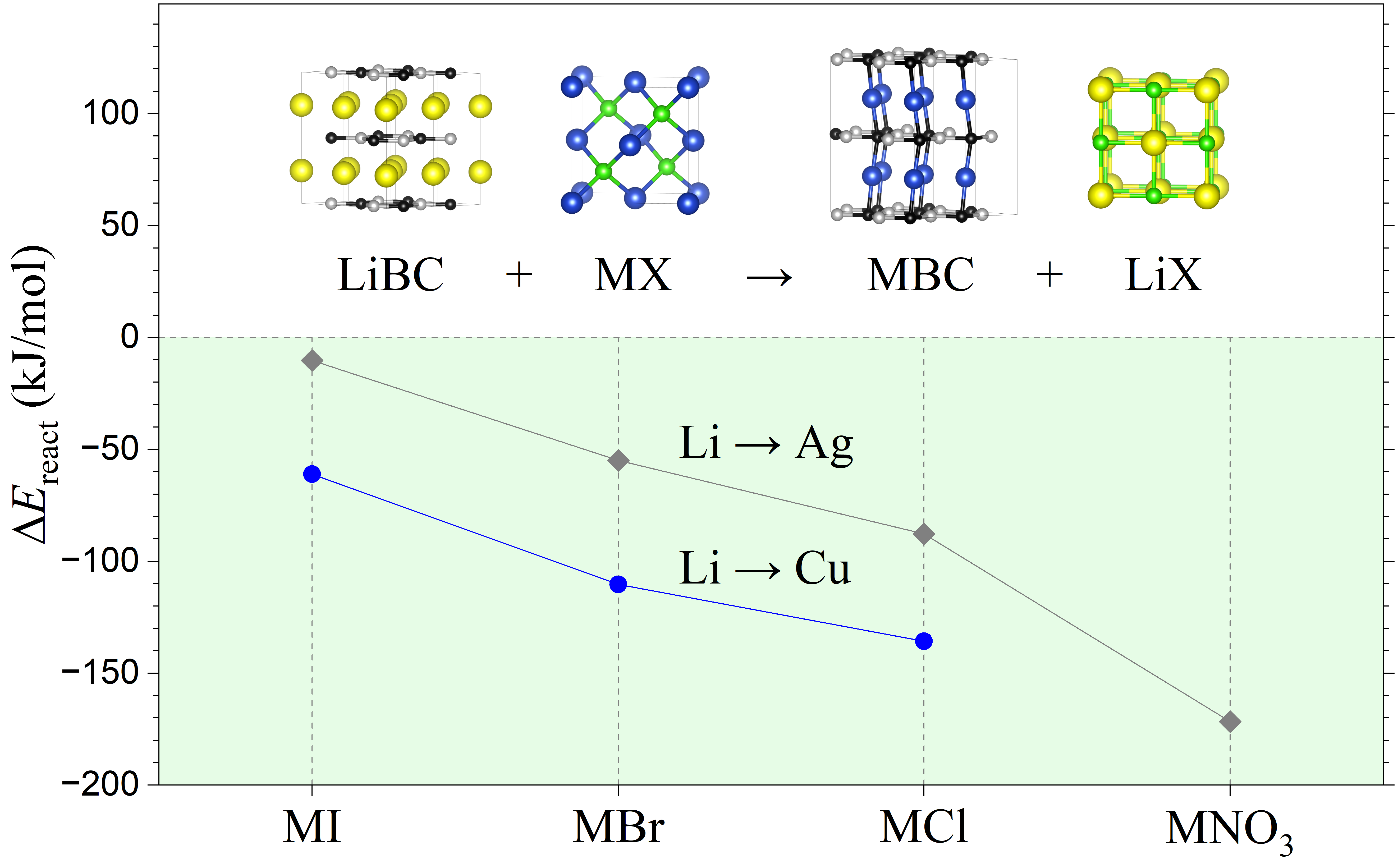}
    \caption{\label{fig-00} Energy per mole for the topochemical reactions between stoichiometric LiBC and metal salts XM (X = I, Br, Cl, or NO$_3$) resulting in the formation of MBC (M = Cu or Ag). Negative energies indicate exothermic reactions.}
\end{figure}
\section{Results and discussion}

\label{sec:Energetics}
\subsection{Topochemical reaction energetics}

Unlike conventional reactions that drive materials toward their most thermodynamically stable phases, topochemical transformations yield metastable products that are most favorable within the kinetically constrained space defined by precursors. The success of topochemical ion exchange in several classes of layered materials suggests that a similar process should be viable for metal borocarbides with strong honeycomb frameworks provided that the net Gibbs free energy change is negative. For example, a partial replacement of Li with another metal induced by reacting LiBC with the corresponding metal halide is described by
\begin{equation}
\text{LiBC} + y\text{MX} \rightarrow \text{Li}_{1-y}\text{M}_y\text{BC} + y\text{LiX}, \nonumber
\end{equation}
where X  is Cl, Br, or I, and M is Na, Cu, or Ag. Suzuki et al.~\cite{Suzuki2024} observed full ion exchange for all reactions with DFT $\Delta E_{\rm react}<-7.3$ kJ/mol ($-0.025$ eV/atom, see Supplementary Note I for unit conversion) and attributed the formation of disordered (Ag,Na)GaO$_2$ alloys with the parent structure to the $-TS_{\text{conf}}$ configurational entropy contribution.

\begin{table}[!t]
\caption{Calculated topochemical reaction energies in kJ per mole for different precursors and ion sources. The energetics of ion swaps in metal iridate~\cite{Abramchuk2017} and gallate~\cite{Suzuki2024} compounds serve as a reference.} \label{tab:T1}
\setlength\tabcolsep{0pt} 
\smallskip 
\begin{tabular*}{\columnwidth}{@{\extracolsep{\fill}} c c  r r r r r}
\hline\hline \noalign{\vskip 1mm}
 Precursor  & Derived  & \multicolumn{5}{c}{$\Delta E_{\rm react}$ (kJ/mol)}   \\
  compound  & compound & M & MI & MBr & MCl & MNO$_3$\\
\hline
\noalign{\vskip 1mm}
LiBC           &   NaBC         &   $+100$ &   $+111$  &   $+104$  &   $+96$ &  $+80$\hspace{0.7em} \\
LiBC           &   CuBC         &   $+158$ &   $-61$  &   $-110$  &   $-136$ &   \\
LiBC           &   AgBC         &   $+197$ &   $-10$  &   $-55$  &   $-88$ &  $-172$\hspace{0.7em} \\
\noalign{\vskip 2mm} 
BeB$_2$C$_2$  &   ZnB$_2$C$_2$  &   $+116$ &   $+114$  &   $+209$  &   $+21$ &   \\
MgB$_2$C$_2$  &   ZnB$_2$C$_2$  &   $+142$ &   $-14$  &   $-58$  &   $-91$ &  $-185$\hspace{0.7em} \\
\noalign{\vskip 2mm} 
Na$_2$IrO$_3$ &   Cu$_2$IrO$_3$ &   $+214$ &   $-17$  &   $-59$  &   $-77$ &   \\
LiGaO$_2$     &   AgGaO$_2$     &   $+165$ &   $+142$  &   $+98$  &   $+65$ &  $-19$\hspace{0.7em} \\
\noalign{\vskip 1mm}
\hline\hline
\end{tabular*}
\end{table}
In the present DFT evaluation of reaction energies, we used structures with honeycomb BC layers and the most stable metal arrangements found in extensive searches detailed in Section~\ref{sec:Stability}. The most favorable mP6-CuBC derivative, for example, turned out to be 45 kJ/mol ($0.157$ eV/atom) lower in energy than the starting hP6-LiBC prototype. The energy difference is significantly higher than the free energy corrections from configurational and vibrational entropy discussed in Section~\ref{sec:Stability}, which highlights the importance of exploring the constrained space. The ground state structures for the known materials, taken from the Materials Project~\cite{MP}, are listed in the Supplementary Material.

The DFT $\Delta E_{\rm react}$ values summarized in Fig.~\ref{fig-00} and Table~\ref{tab:T1} indicate that full ion exchange is a downhill reaction for several precursor-salt combinations. The monovalent Li could be replaced with either Cu or Ag introduced as any of the considered halides. The Li-to-Cu ion exchange should be easier to achieve kinetically due to their similar sizes and a moderate 13\% expansion of the interlayer spacing in the predicted mP6-CuBC. The substantial 32\% interlayer expansion needed to accommodate Ag in the predicted hP3-AgBC and the less favorable energetics, by $\sim 50$~kJ/mol relative to CuBC for each salt, may hinder the reaction rate. $\Delta E_{\rm react} = -10$~kJ/mol for the least exothermic LiBC + AgI is comparable to values associated with the formation of mixed-metal products~\cite{Suzuki2024}. Our calculations show that the thermodynamics could be improved significantly, to $-172$~kJ/mol, by switching to an alternative AgNO$_3$ ion source, \DG{but limiting the energy release by using AgI/AgBr mixtures with optimized ratios could increase the chances of preserving the BC morphology}.

Among the known alkaline-earth metal borocarbides, BeB$_2$C$_2$ has an oP20 ground state with differently stacked BC honeycomb layers~\cite{Hofmann2008}, MgB$_2$C$_2$ crystallizes in a complex oS80 structure with BC honeycomb layers sandwiching Mg rhombus-shaped patches~\cite{Worle1994}, and CaB$_2$C$_2$ adopts a tI20 structure with 4-8-polygon BC layers~\cite{Albert1999}. Our global structure searches indicate that ZnB$_2$C$_2$ would be isostructural to BeB$_2$C$_2$ and could be synthesized via topochemical Mg-to-Zn exchange leading to a minor 2\% interlayer spacing reduction. Reactions with the proposed NaBC precursor, shown in Supplementary Table~I, are even more exothermic. Given the promise of this material, we revisited the ternary system by considering two other stable Na-B-C phases recently entered into the Materials Project database, NaB$_{13}$C$_2$ (mp-3293300), synthesized under pressure~\cite{Yuan2022}, and NaBC$_7$ (mp-3230146), a possible Na-intercalated B-doped graphite. According to our revised relative stability plots (Supplementary Fig.~1), NaBC remains a low-temperature ground state and, as such, may not be accessible through high-temperature synthesis from the elements.

\begin{figure}[t!]
   \centering
\includegraphics[width=0.42\textwidth]{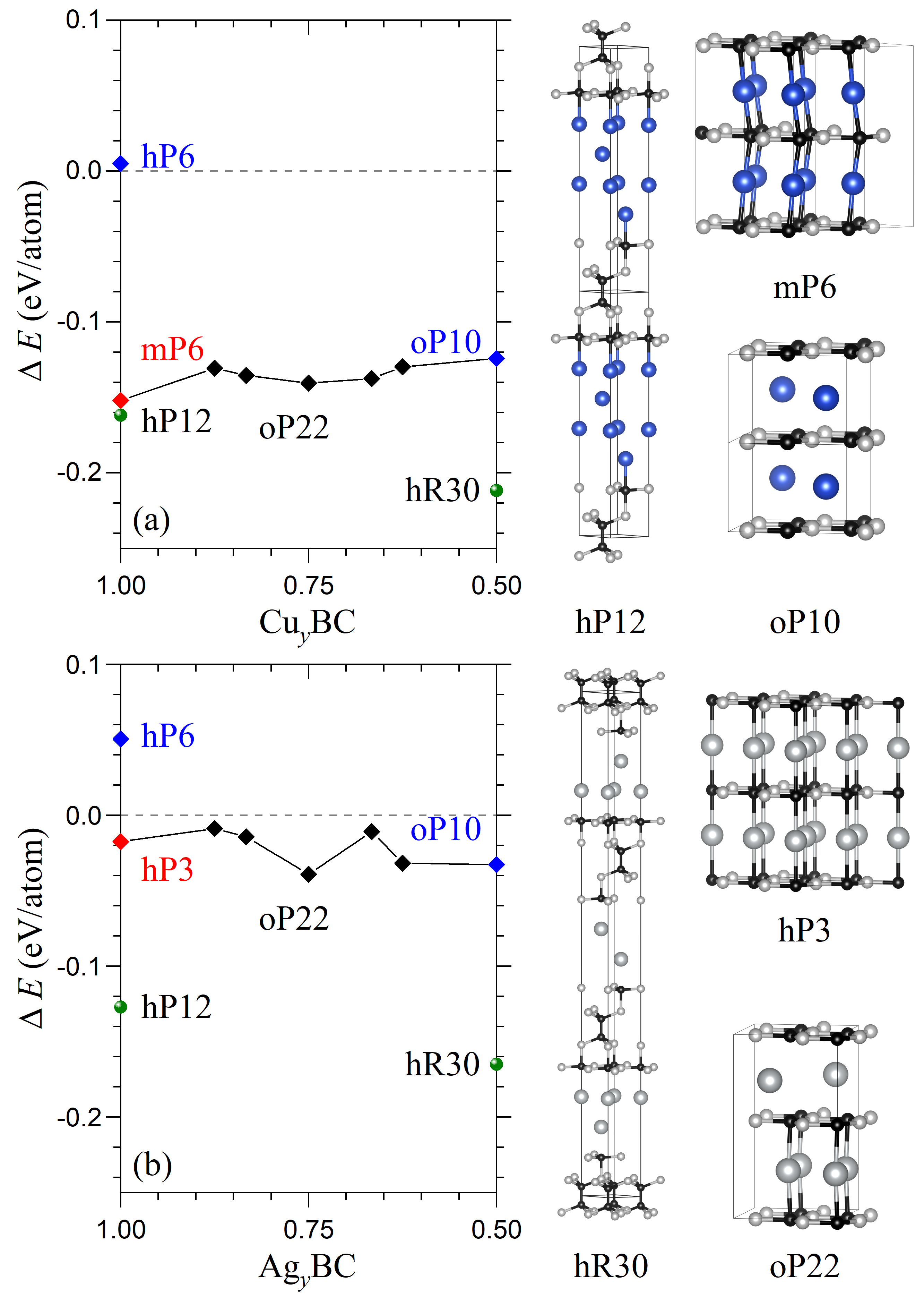}
    \caption{\label{fig-02} Relative energies of Cu$_x$BC and Ag$_x$BC phases referenced to a hypothetical fully deintercalated BC phase and the elemental ground states. The line connects the lowest-energy phases with the honeycomb layer morphology. Both metals prefer to occupy dumbbell sites at $x=1$ (red points) but shift to interstitial sites at $x=1/2$ (blue points). The green points correspond to alternative motifs with bridged layers. The B, C, Cu, and Ag atoms in the displayed crystal structures are shown with light gray, black, blue, and gray spheres, respectively.}
\end{figure}

To illustrate the level of reliability of the DFT-based predictions for metal borocarbides, Table~\ref{tab:T1} lists calculated $\Delta E_{\rm react}$ for select metal oxides investigated in previous studies. The large negative $-77$~kJ/mol is consistent with the successful Na-to-Cu swap via Na$_2$IrO$_3$ + CuCl~\cite{Abramchuk2017}. Our calculated energy changes for the Li-to-Ag swap in delafossites agree within $\sim 5$~kJ/mol relative to reported values obtained with the PBEsol functional~\cite{Suzuki2024} and correctly identify AgNO$_3$ as the only salt enabling synthesis of AgGaO$_2$. Notably, the calculated volume change in this topotactic reaction is fairly isotropic and reaches 26\%. 

Topochemical reactions discussed so far involve stoichiometric precursors but the demonstrated~\cite{Bharathi2002, Zhao2003,Fogg2003a,Fogg2003b,Fogg2006, Kalkan2019} or predicted~\cite{Tomassetti2024a,Tomassetti2024b} thermal deintercalation of metal borocarbides offers a new dimension in the design of derived materials. Namely, while the direct replacement of the intercalant (Li, Be, or Mg) with a new metal (Na, Cu, Ag, or Zn) is disfavored for all known metal borocarbides listed in Table ~\ref{tab:T1}, reintercalation of initially delithiated Li$_x$BC with different metals (Na, K, Mg, or Ca) has been shown to be thermodynamically favorable~\cite{Tomassetti2024a}. Hence, our present study also includes the screening of non-stoichiometric ternary M$_{x=y}$BC and mixed-metal quaternary Li$_{x-y}$M$_y$BC derivatives with M = Cu or Ag.

\subsection{Thermodynamic stability} 
\label{sec:Stability}

The exploration of the constrained Li$_{x-y}$M$_y$BC ($1\ge x\ge 1/2$ and $1 > x-y \ge 0$) phase space originating from the precursor composition relied on two complementary strategies. Combinatorial screening of supercells with honeycomb BC layers enabled identification of stable intercalant decorations, while global evolutionary searches helped establish favorable composition-dependent motifs beyond the parent morphology. 

\begin{figure}[t!]
   \centering
\includegraphics[width=0.45\textwidth]{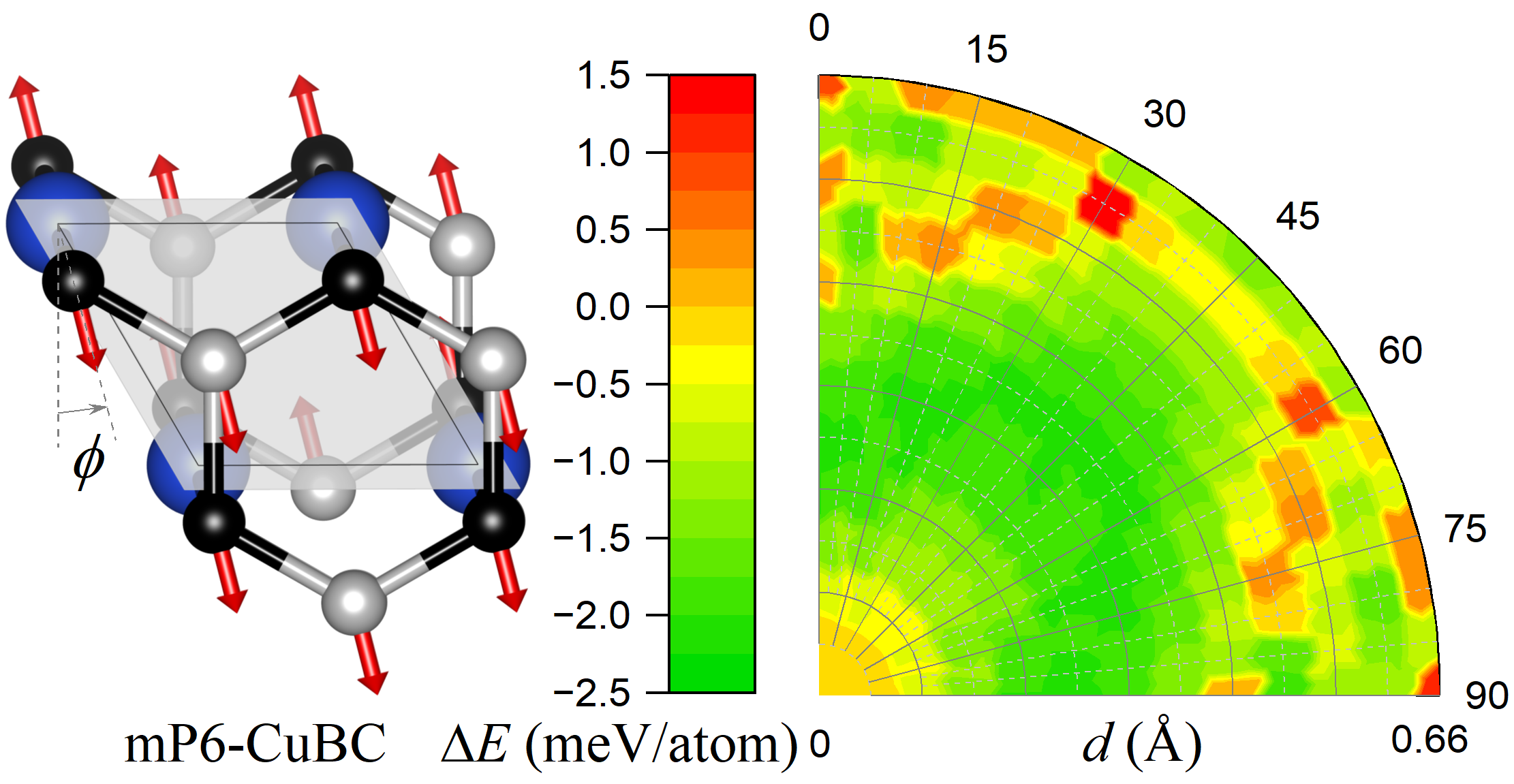}
    \caption{\label{fig-03} Change in the CuBC energy as a function of interlayer shift angle ($\phi)$ and magnitude ($d$) away from the hP3 configuration. The CuBC derivatives with the lowest mP6 symmetry were obtained by displacing adjacent BC layers in the $1\times1\times2$ hP3 supercell in the opposite directions.}
\end{figure}

Our first evolutionary runs for ternary M$_y$BC compounds with M = Cu or Ag uncovered a notable difference between preferred locations of the transition and alkali or alkaline-earth metals in the BC honeycomb matrices. In contrast to Li, Na, K, Mg, Ca, or their Li-M combinations that tend to be in or near the middle of the hexagons between AA$^\prime$-stacked BC layers~\cite{Tomassetti2024a,Tomassetti2024b}, Cu and Ag showed a clear preference to form C-M-C dumbbells forcing the AA stacking sequence. Therefore, we generalized the decoration sampling protocol to allow initial positions of the transition metals to be either at the interstitial or dumbbell sites and imposed a constraint on the minimum distance between neighboring metals to avoid unphysical starting decorations. The adjustment expanded the number of unique states by over 1000 structures across the considered composition range. Figure~\ref{fig-02} displays the lowest-energy phases at each composition for dumbbell, interstitial, or mixed decorations. The results show that the two ternary systems prefer to adopt dumbbell and interstitial configurations at the full and half-filled levels, respectively. Phases at intermediate stoichiometries, e.g., oP22 at $y=3/4$, represent mixtures of fully-filled galleries with dumbbells and half-filled galleries with interstitials.
\begin{figure}[t!]
   \centering
\includegraphics[width=0.45\textwidth]{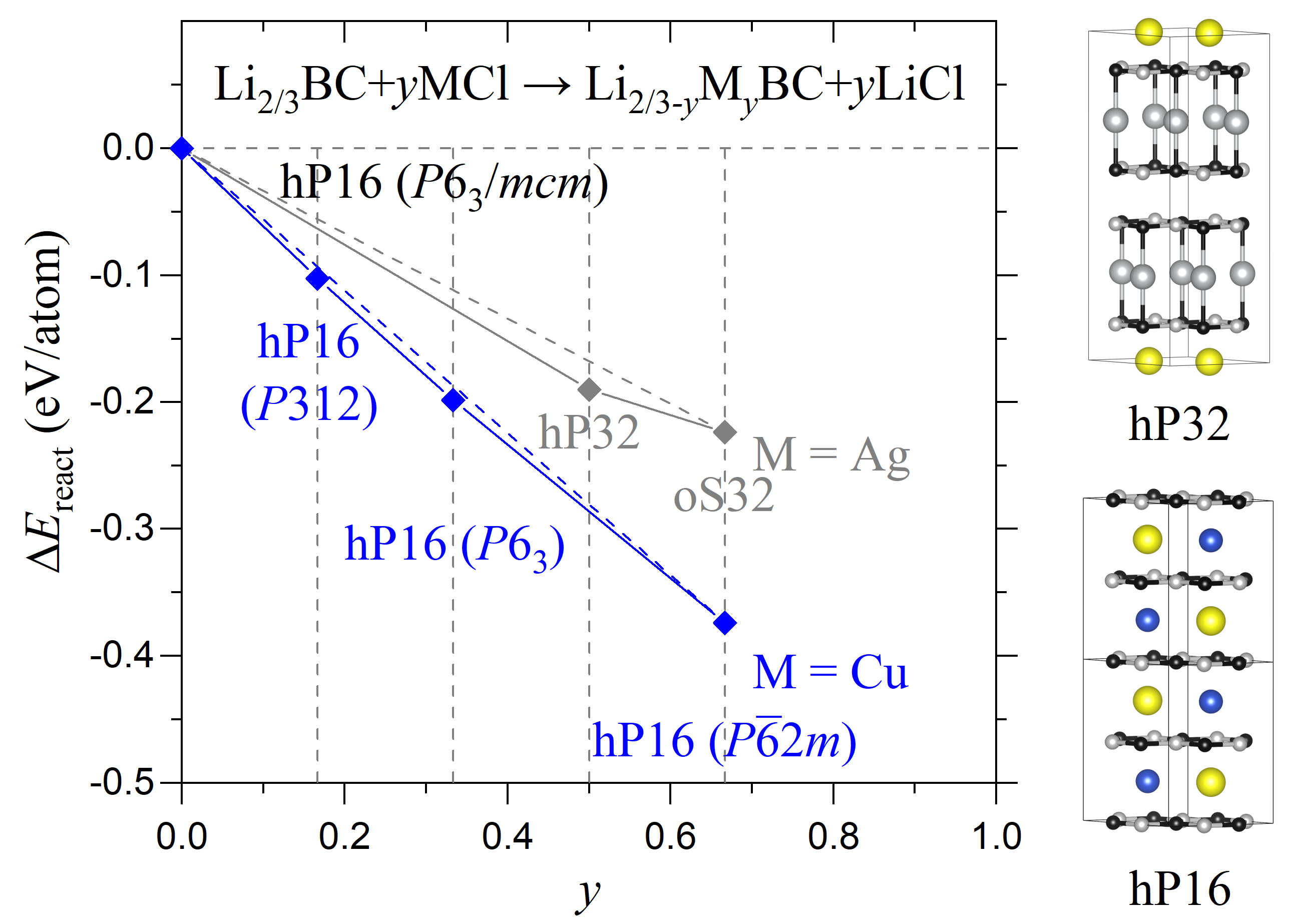}
    \caption{\label{fig-04} Reaction energy per atom of topochemical ion exchange between 
    Li$_{2/3}$BC and $y$MCl (M = Cu or Ag). The solid points mark quaternary phases stable with respect to Li$_{2/3}$BC and M$_{2/3}$BC and forming the local convex hull, such as the displayed hP16-Li$_{1/3}$Cu$_{1/3}$BC ($P6_3$) and hP32-Li$_{1/6}$Cu$_{1/2}$BC.}
\end{figure}
The stoichiometric CuBC and AgBC, natural candidates to form with the LiBC precursor, warrant closer examination. AgBC is dynamically and thermally stable in the hP3 structure with linear C–Ag–C links, whereas hP3-CuBC exhibits imaginary phonon modes shearing the BC layers (see Supplementary Figs.~2 and~3 and Section~\ref{sec:eph}). These observations align with the known tendency of both metals to form dumbbells, and with the electronic frustration observed in Cu-based materials, as Cu$^{1+}$ often disproportionates into Cu$^{0}$ and Cu$^{2+}$~\cite{Zhang2024}. Displacement of the atoms along the A-point imaginary phonon eigenmode in the doubled unit cell generated a dynamically stable mP6-CuBC phase 2.5 meV/atom lower in energy (Fig.~\ref{fig-03}). However, the presence of the radially degenerate minima around the hP3 stationary point shows that CuBC may have a complex structure with stacking disorder that will require further investigation. For example, the r$^2$SCAN+rVV10 functional, better suited to handle variable oxidation states and capture medium-range correlation~\cite{Zhang2024}, also produces an imaginary phonon mode at the A point but favors mP6 over hP3 by only 0.5 meV/atom. In general, predicting and resolving distorted configurations represent both computational and experimental challenges, with some of the proposed revisions, such as the lower-symmetry ground state of Cu$_2$IrO$_3$, have yet to be confirmed~\cite{ak44}. The electronic structure analysis in Section~\ref{sec:eph} sheds light on the origin and consequences of the dumbbell linkage.

\begin{figure}[t!]
   \centering
\includegraphics[width=0.45\textwidth]{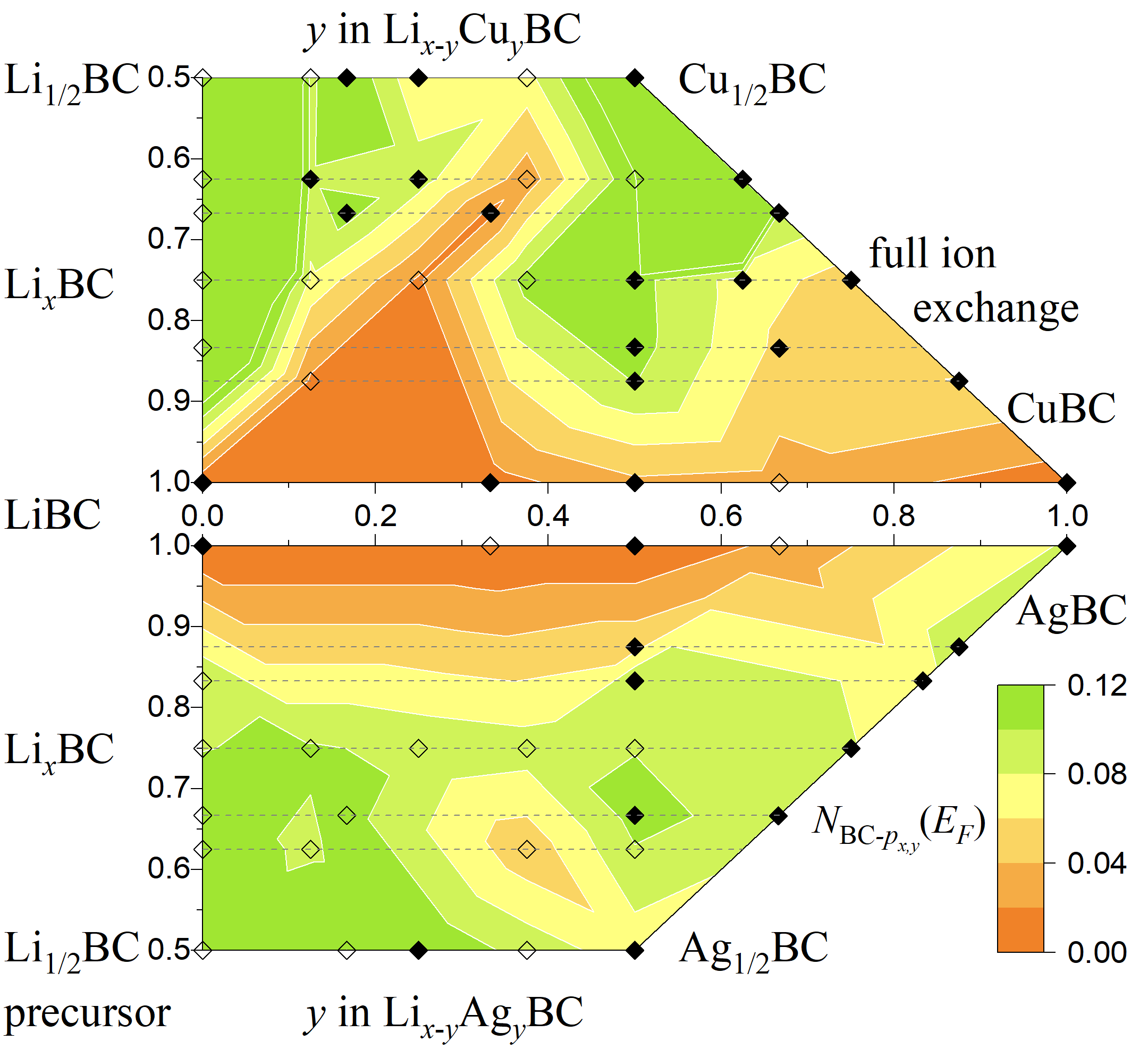}
    \caption{\label{fig-05} Projected DOS from BC-$p_{x,y}$ (states/(eV atom)) in  Li$_{x-y}$M$_y$BC compounds with M = Cu or Ag. The solid diamonds mark ternary and locally stable quaternary compounds at 600~K. The contour map is based on compounds within 25 meV/atom to the local convex hulls marked with solid or hollow diamonds.}
\end{figure}

\begin{figure*}[!t]
   \centering
\includegraphics[width=1.0\textwidth]{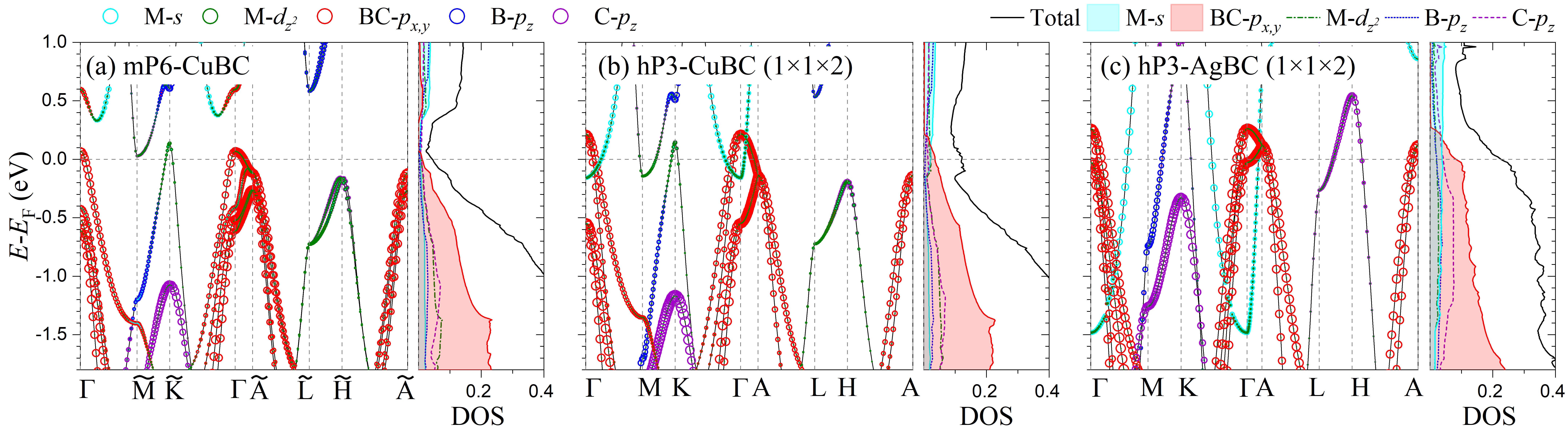}
    \caption{\label{fig-06} Orbital-resolved electronic band structure and DOS in states/(eV atom) in CuBC and AgBC phases with straight (hP3) and tilted (mP6) C-M-C dumbbells. For direct comparison, the hP3 unit cells are doubled along the $c$ axis, and the high-symmetry points along the standard path for mP6 are labeled using hexagonal notation (\~{M}=Z, \~{K}=$(1/3,1/3,0)$, \~{A}=Y, \~{L}=D, \~{H}=$(1/3,1/3,1/2)$). The $p_{x,y}$ characters of B and C are combined, and only the $d_{z^2}$ character of the $d$ manifold is shown for Cu or Ag, since the combined contribution from all other $d$-states is five times smaller at the Fermi level. The key feature defining the hole-doping level of the BC-$p_{x,y}$ states is the position of the M-$s$ band edge: (a) above $E_{\text{F}}$; (b) $-0.15$ eV below $E_{\text{F}}$; and (c) $-1.5$ eV below $E_{\text{F}}$.}
\end{figure*}

Figure~\ref{fig-02} demonstrates that insertion of the transition metals into a hypothetical empty honeycomb BC framework leads to significantly lower stabilization, below $0.15$ eV/atom for CuBC and $0.01$ eV/atom for AgBC, compared to $\sim 0.4-0.7$ eV/atom for LiBC, NaBC, or MgB$_2$C$_2$~\cite{Kharabadze2023,Tomassetti2024b}. The metastable layered transition metal borocarbides are also further above the global convex hull facet of M, C, and B$_4$C, by at least $0.16$ eV/atom for Cu$_y$BC and $0.3$ eV/atom for Ag$_y$BC in the full investigated range of $y$. \DG{In fact, CuBC and AgBC have positive formation energies of $0.12$ and $0.25$ eV/atom, respectively, but the unfavorable energetics prohibiting the direct synthesis from the elements does not automatically disallow exothermic reactions steered by kinetics. For example, a related layered BC$_3$, unstable with respect to B and C by a comparable $0.20$ eV/atom~\cite{Kharabadze2023}, has been synthesized from different precursors~\cite{Bhaskar2021, king2015}. A more pertinent question is whether M$_y$BC compounds have polymorphs with derived morphologies, such as linked-layer structures favored in significantly delithiated LiBC~\cite{Tomassetti2024a}, that could form instead or serve as transient points {\it en route} to full decomposition.} Our evolutionary searches for $y=1$ and $y=1/2$ indicate that more stable hP12 ($P3m1$) and hP30 configurations with metal blocks separated from bridged BC frameworks exist even \DG{in the stochiometric CuBC and AgBC}. To illustrate that this is not an uncommon situation, we performed evolutionary searches for AgGaO$_2$ and found an hP12 ($R\overline{3}m$) structure with an alternative morphology 0.12 eV/atom below the synthesized wurtzite prototype (see Supplementary Fig.~4). Therefore, one can expect that no major rebonding will occur in the topochemical Li-to-Ag ion exchange for borocarbides either. 

To identify possible products of partial ion exchange in stoichiometric and delithiated Li$_x$BC precursors, we constructed supercells with $x$ = 1, 7/8, 5/6, 3/4, 2/3, 5/8, and 1/2 and screened over 3600 layered Li$_{x-y}$M$_y$BC configurations for each transition metal. Representative relative stabilities of quaternary compounds are shown in Fig.~\ref{fig-04} for $x=2/3$, with results for other compositions complied in Supplementary Fig.~5 and the locally stable phases at 600 K collected in Fig.~\ref{fig-05}.

The results show that ordered mixed-metal derivatives with Cu and Ag below the Li$_x$BC and M$_x$BC line are present at each starting composition $x$. The two structures displayed in Fig.~\ref{fig-04} reflect the observed preferences for Cu to co-exist with Li within galleries and for Ag to avoid mixing. 13 Cu-based and 7 Ag-based quaternary compounds are locally stable, by $\Delta E_{\rm rel}$ between $-0.004$ and $-0.040$ eV/atom at $T=0$ K. The vibrational entropy contribution at 600 K moves the reaction free energies upward by an average of $0.014$ eV/atom and disfavors some of the 
quaternary phases by up to $0.011$ eV/atom, fully destabilizing two of them. For comparison, the configuration entropy stabilization estimated at $T=600$~K amounts to only $-0.012$ eV/atom (see Supplementary Note II). These contributions are considerably smaller in magnitude than the reaction energies of full ion exchange of about $-0.4$ eV/atom. In light of the previously predicted and observed transformations~\cite{Suzuki2024}, the topochemical reactions are expected to proceed until full Li to Cu or Ag ion exchange. If quaternary compounds do form, they may have disordered double-metal arrangements, unless the samples can be annealed.

The Li occupancy pattern in partially delithiated Li$_x$BC precursors may also play a critical role in defining the reaction products. While staging models were proposed to explain lithium-deficient borocarbides~\cite{Kalkan2019}, the lowest-energy configurations in our DFT calculations at all compositions featured uniform distributions of Li along the $c$ axis ~\cite{Tomassetti2024a}. Therefore, the most stable identified Li$_{x-y}$M$_y$BC derivatives with uneven metal content per layer, such as hP32-Li$_{1/6}$Cu$_{1/2}$BC with fully Ag-filled and metal-free galleries (Fig.~\ref{fig-04}) as part of the most viable series with $y=1/2$ for both transition metals  (Fig.~\ref{fig-05}), might not be reachable kinetically. Nevertheless, the representative models with relatively small unit cells are useful for investigating various properties of the mixed-metal borocarbides. In particular, the 60 meV/\AA$^2$ binding energy per area between BC-Ag-BC-Li$_{2/3}$-BC-Ag-BC blocks makes the hP32 quaternary phase potentially exfoliable~\cite{bjorkman2012,Campi2023}.

\subsection{Electronic and superconducting properties}
\label{sec:eph}

Stoichiometric layered borocarbides of alkali and alkaline-earth metals exhibit semiconducting behavior expected for materials satisfying the 8-electron rule. Early studies with standard semi-local functionals found similar band gaps of $\sim 1$ eV in LiBC~\cite{Rosner2002}, MgB$_2$C$_2$~\cite{Verma2003}, and BeB$_2$C$_2$~\cite{Hofmann2008}. The band gaps in our HSE06 (optB86b-vdW) calculations are 1.61 (0.84) eV in LiBC, 1.98 (1.18) eV in MgB$_2$C$_2$, 1.29 (0.39) eV in BeB$_2$C$_2$, 1.87 (0.95) eV in NaBC, and 1.67 (0.95) eV in ZnB$_2$C$_2$. The findings underscore yet again the uniqueness of the isoelectronic but intrinsically hole-doped MgB$_2$ thermodynamically stable under ambient conditions.

\begin{table}[!t]
\caption{\CT{Projected BC-${p_{x,y}}$ density of states at the Fermi level, $N_{p_{x,y}}(E_{\text{F}})$ (states/(eV atom)), logarithmic average phonon frequency, $\omega_{\rm log}$ (meV), and superconducting critical temperature, $T_{\rm c}$, calculated using the Allen-Dynes formula~\cite{AD} with $\mu^*=0.1$, the iME formalism with $\mu^*=0.1$ (e-ph matrix elements computed with \textsc{Quantum} ESPRESSO on coarse grids), or the aME formalism with $\mu^*=0.2$ (e-ph matrix elements computed with EPW on fine grids) for select metal borocarbides. For comparison, we list MgB$_2$ properties evaluated with similar settings in our previous study~\cite{Kafle2022}. Among the considered quaternary phases, only Li$_{1/2}$Ag$_{1/4}$BC is above the local convex hull (by 12 meV/atom).}} \label{tab:T2}
\setlength\tabcolsep{0pt} 
\smallskip 
\begin{tabular*}{\columnwidth}{@{\extracolsep{\fill}} l c c c c c c c c c}
\hline\hline \noalign{\vskip 1mm}
 Composition  &  Pearson & $N_{p_{x,y}}$  & \CT{$\omega_{\rm log}$}  & \CT{AD} &  iME  & aME  \\
              &  symbol  &  at $E_{\text{F}}$ & \CT{(meV)}  & \CT{$T_{\rm c}$(K)}      & $T_{\rm c}$(K)  & $T_{\rm c}$(K) \\
\hline
\noalign{\vskip 1mm}
\CT{MgB$_2$}                &   \CT{hP3}  &  \AK{0.098}  &   \CT{60.3}  & \CT{18.1} &  \CT{24.9}   &  \CT{41} \\\noalign{\vskip 2mm}
CuBC                   &   mP6  &  0.009  &  \CT{19.9}  & \CT{0.3} &  0.4   & \\
Cu$_{2/3}$BC           &   hP16 &  0.097  &  \CT{34.9}  & \CT{14.7} &  16.0  & \\
Cu$_{5/8}$BC           &   hP21 &  0.119  &  \CT{42.8}  & \CT{23.4}  &  26.6  &\\
\noalign{\vskip 2mm} 
AgBC                   &   hP3  &  0.083   &  \CT{50.5}  & \CT{10.6} &  12.0  & 56  \\
Ag$_{1/2}$BC           &   oP10 &  0.069   &  \CT{20.1}  & \CT{7.0}  &  6.5  & \\
\noalign{\vskip 2mm} 
Li$_{1/6}$Cu$_{2/3}$BC &  hP17 &  0.055 &   \CT{35.5}   &  \CT{2.2}  &  2.5  & \\  
Li$_{1/2}$Cu$_{1/6}$BC &  hP16 &  0.108   &   \CT{63.0}  & \CT{11.3}  &  12.9 &  \\
\noalign{\vskip 2mm} 
Li$_{1/2}$Ag$_{1/4}$BC &  hP22 &  0.095  &   \CT{57.7}   &   \CT{10.9}  &  12.7 &  \\ 
Li$_{1/4}$Ag$_{1/2}$BC &  oI22 &  0.099   &   \CT{55.5}   & \CT{15.7}  &   18.2  & \\ 
\noalign{\vskip 1mm}
\hline\hline
\end{tabular*}
\end{table}

\begin{figure*}[t!]
   \centering
\includegraphics[width=0.99\textwidth]{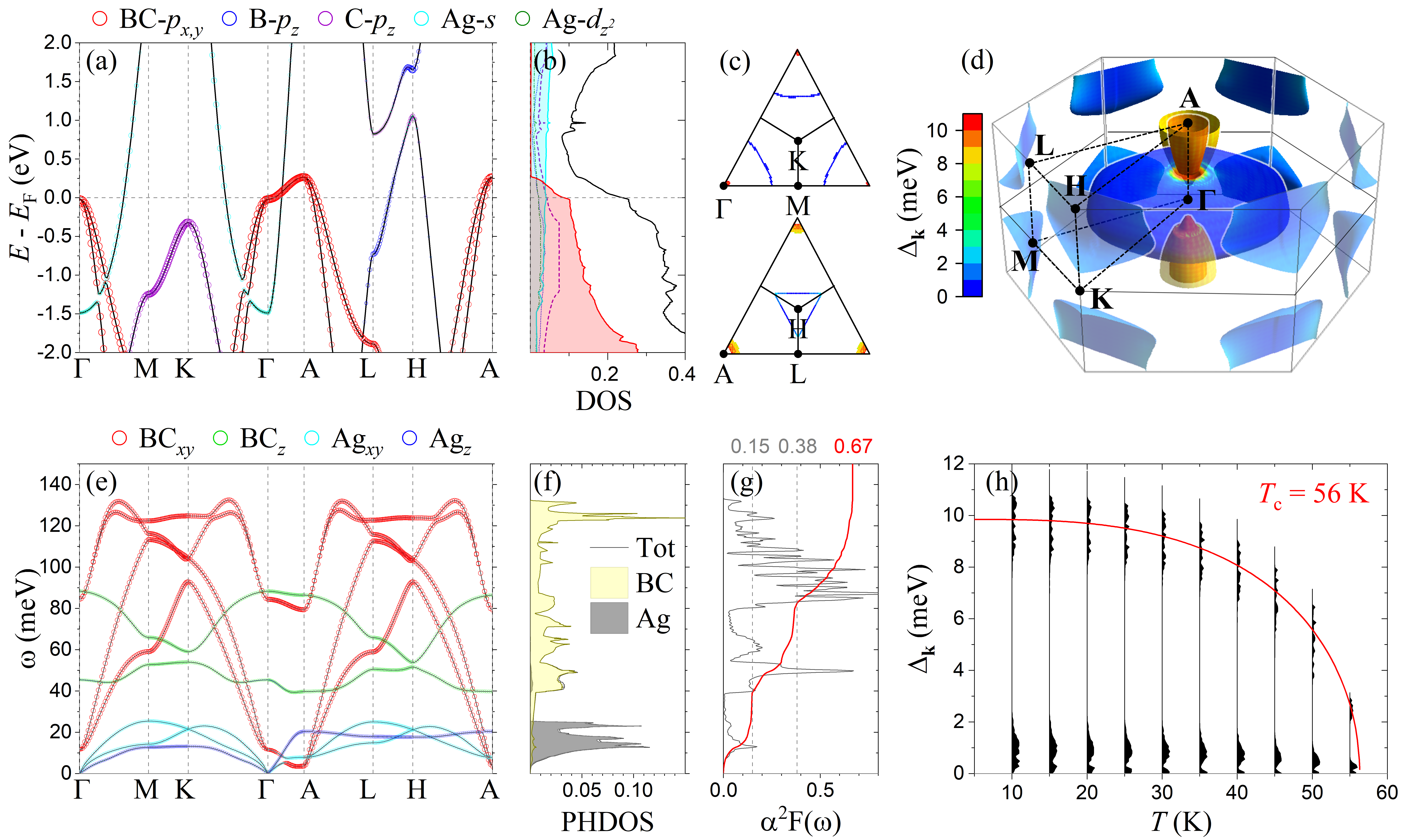}
    \caption{\label{fig-07} Electronic, vibrational, and superconducting properties of hP3-AgBC. (a) Orbital-resolved band structure. (b) Total and projected DOS in states/(eV atom). (c) Fermi surface cuts in the $\Gamma$-M-K and A-L-H planes. (d) Fermi surfaces with the color map showing superconducting gap values. (e) Phonon dispersion with atom- and direction-resolved eigenmodes. (f) Phonon DOS in states/(meV atom) projected on BC and Ag. (g) Eliashberg spectral function and e-ph coupling strength integrated to 0.67. (h) Superconducting gap as a function of temperature extrapolated to vanish at $T_{\rm c} = 56$~K for $\mu^*=0.2$.}
\end{figure*}

Replacing Li with Cu or Ag turns the stoichiometric borocarbides into metals and potential MgB$_2$-type superconductors. To disentangle the dependence of the electronic states on the structural and chemical factors, we display projections of the most relevant states around the Fermi level for three layered phases in Fig.~\ref{fig-06}. Due to the strong mixing of B and C in-plane orbitals, we group their contributions as BC-$p_{x,y}$  character. The interlayer bridging by the metal-carbon dumbbells makes the M-$d_{z^2}$ states hybridize with the C-$p_z$, B-$p_z$, and M-$s$ states near the Fermi level in CuBC. A key feature in both hP3 phases is the partial occupation of the nearly-free electron M-$s$ band, which contributes significantly to the total DOS at the Fermi level and results in hole doping of the BC-$p_{x,y}$ states. In the mP6-CuBC derivative with only 0.5~\% shorter interlayer spacing, the M-$s$ states are no longer filled and the in-plane BC states are barely hole doped, which leads to the emergence of a pseudogap. The much deeper position of the M-$s$ band edge in hP3-AgBC, at $-1.5$ eV at $\Gamma$, prevents the material from emptying the nearly free electron states and reducing the DOS at the Fermi level via a similar structural distortion. Additional tests in Supplementary Fig.~6 reveal that the size difference between the two transition metals, that determines the interlayer distance, is the primary factor for the substantially different electronic landscapes in the two layered stoichiometric materials. While AgBC is dynamically stable at 0 GPa, both CuBC and AgBC are (further) destabilized under pressure (see Supplementary Fig.~7).

The projected DOS map in Fig.~\ref{fig-05} shows that AgBC is the only viable considered compound with substantially hole-doped BC-$p_{x,y}$ states in the fully intercalated Li$_{1-y}$M$_y$BC series. In order to achieve similar doping levels of the covalent framework in non-stoichiometric borocarbides, one has to start with sufficiently delithiated Li$_x$BC precursors. Namely, the first locally stable phases with at least 0.08 states/(eV atom) BC-$p_{x,y}$ contribution appear at $x=7/8$ and $x=3/4$ in the Cu- and Ag-based quaternaries, respectively. The low projected DOS values in some mixed-metal compounds with depleted metal content, e.g., in Li$_{1/3}$Cu$_{1/3}$BC, indicate a complex interplay between competing states at the Fermi level. In fact, metal filling reduction appears necessary for the onset of superconductivity in Cu$_y$BC ($y=2/3$) but would be detrimental for the $T_{\text{c}}$ in Ag$_y$BC.

The following superconductivity analysis, based on isotropic Migdal-Eliashberg (iME) estimates of $T_{\text{c}}$, supports these qualitative observations. The results listed in Table~\ref{tab:T2} and Supplementary Fig.~8 show that CuBC has negligible $T_{\text{c}}$ at $y=1$ but becomes a promising superconductor at lower metal concentrations, while AgBC has a higher $T_{\text{c}}$ estimate than the half-full counterpart. The phonon dispersions in Supplementary Fig.~9 show that minor dynamical instabilities, common in layered materials~\cite{Mishra2024}, persist in some of the identified phases and would need to be carefully addressed, but the presented Eliashberg spectral functions indicate that these materials should have strong e-ph coupling~\cite{nepal2024}. Overall, several ternary and quaternary borocarbides with the transition metals have $T_{\text{c}}$ between 10 and 23~K, comparable to the values in Mg-Li-B~\cite{Kafle2022}, Li-M-BC~\cite{Tomassetti2024a}, and Na(Mg)-BC~\cite{Tomassetti2024b} obtained at the same level of theory. Since these studies have shown a general underestimation of the isotropic $T_{\text{c}}$ by a factor of two to four in such layered materials, we have performed aME calculations for the representative AgBC material.

The results in Fig.~\ref{fig-07} establish hP3-AgBC as a compelling superconductor that merges key features of MgB$_2$~\cite{An2001,Kortus2001} and graphite intercalation compounds (GICs)~\cite{Calandra2005,Mazin2005}, yet deviates from their prototypical behavior. The characteristic hourglass-shaped Fermi surfaces from heavily hole-doped B-$p_{x,y}$ states in MgB$_2$~\cite{Kortus2001,Kafle2022} or essentially cylindrical sheets in LiB~\cite{Liu2007,Calandra2007,Kafle2022} are replaced by elongated ellipsoids in AgBC, a consequence of an insufficient doping of the BC-$p_{x,y}$ bands that now cross the Fermi level between $\Gamma$ and A. The BC-$p_z$ bands generate closed sheets around the H point with the triangular cross-section in the A-H-L plane. In contrast to the electronically inert Mg$^{2+}$, Ag retains some charge on its $s$ orbitals that generate Fermi surfaces enclosing the $\Gamma$ point, but the pronounced $c$-axis anisotropy deforms the NFE-type spheres observed in CaC$_6$~\cite{Sanna2007} into pancake shapes. 

The existence of the three large Fermi surface pockets defines three sizable contributions to the e-ph coupling ($\lambda$) that can be distinguished by examining the phonon dispersion and  Eliashberg spectral function displayed in Fig.~\ref{fig-07}(e-g). In contrast to MgB$_2$ with the same $\lambda =0.67$~\cite{Kafle2022}, only about 0.30 (40\%) arises from the in-plane bond stretching mode $E^{\prime}$ coupled with the BC-$p_{x,y}$ states, manifested through the pronounced $\sim 30$ \% phonon softening down to 85 meV at $\Gamma$ and the familiar splitting of these electronic states under the frozen phonon displacements (Supplementary Fig.~10). The rest comes from BC$_{z}$ phonons in the 40-60 meV range, particularly along the M-K and L-H directions, and soft mixed modes around 10-20 meV as further evidenced by the mode and $\bq$-point resolved $\lambda$ (see Supplementary Fig.~11). The in-plane intercalant phonons play a dominant role in CaC$_6$~\cite{Calandra2005} but not in AgBC, where Ag is linked with C in dumbbells. 
 
To estimate the superconducting $T_{\rm c}$, we solved the anisotropic Migdal-Eliashberg (aME) equations and tracked the evolution of the superconducting gap $\Delta_\bk$ with temperature. As shown in Fig.~\ref{fig-07}(h), AgBC exhibits two-gap superconductivity with pronounced anisotropy, similar to MgB$_2$. The smaller gap lies below 2~meV associated with the pancake-shaped Fermi surface around $\Gamma$ and the closed pocket surrounding H, while the larger has an average around 10~meV, residing on the ellipsoidal surfaces linked to the BC-$p_{x,y}$ states.  Both gaps close around $T=56$~K for a Coulomb pseudopotential $\mu^*=0.2$, chosen such that the experimental $T_{\rm c}$ of MgB$_2$ is reproduced under similar computational settings~\cite{Kafle2022}. \DG{The comparison of logarithmic average phonon frequency and the corresponding Allen-Dynes estimates listed in Table II illustrates that higher $\omega_{\rm log}$ values (60.3 meV in MgB$_2$ versus 50.5 meV in AgBC) do not necessarily mean a higher $T_{\rm c}$ in anisotropic materials with multiple gaps.} Due to the nearly five-fold boost in $T_{\rm c}$, we expect the other iME critical temperature estimates to be increased substantially as well. Lastly, we investigated the effect of doping AgBC by also calculating the aME $T_{\rm c}$ after rigidly shifting the Fermi level by $\pm0.1$~eV. The results in Supplementary Fig.DG{~12} show that moderate electron-doping decreases the $T_{\rm c}$ down to 36~K, stemming from the loss of BC-$p_{x,y}$ states, while hole-doping increases the $T_{\rm c}$ to 61~K.

\section{Discussion and Conclusions}
\label{sec:conclusions}

The presented findings provide further evidence that borocarbides comprising honeycomb layers intercalated with low-valent metals hold great promise as high-$T_{\rm c}$ and potentially practical superconductors~\cite{Tomassetti2024a,Tomassetti2024b}. Such compounds are uniquely suited to possess hole-doped covalent bonds requisite for MgB$_2$-type superconductivity. Indeed, the B-B honeycomb framework is intrinsically two electrons deficient and, once stabilized, resists chemical modifications that further hole-dope the B-$p_{x,y}$ states~\cite{ak14, Canfield2003}. Graphite is already stable and easy to intercalate, but the resulting superconductors with electron-doped NFE states exhibit relatively low $T_{\rm c}$~\cite{Calandra2005,Yang2014}, with potential to exceed 30 K, e.g., in Na-C, under pressure~\cite{Hao2023,Mishra2024}. 
While the B-C framework represents a favorable middle ground between the chemically active B-B and the relatively inert C-C counterparts, the choice of the intercalant remains important: the larger and higher-valent metals like Ca and Sr tend to stabilize clathrate or layered but non-honeycomb structures that lack the key MgB$_2$ features~\cite{Geng2023,Hayami2024}.

The proposed topochemical ion exchange offers some practical advantages over our previously considered soft chemistry reintercalation of layered metal borocarbides~\cite{Kharabadze2023,Tomassetti2024a,Tomassetti2024b}. The favorable thermodynamics of the exchange reaction is ensured not through initial delithiation of LiBC but rather through the introduction of new metals in the form of halide or nitrate salts, which helps avoid the high-temperature deintercalation process and reduce the possibility of structural disorder in the precursor. Replacement of Li with Cu or Ag, successfully achieved in wurtzite-type oxides~\cite{Suzuki2024} with arguably less well-defined layered frameworks shown to be metastable in our study, can be expected to be more natural in the case of honeycomb LiBC. Synthesis of the metastable CuBC or AgBC identified with extensive structure searches would represent the first \textit{theory-guided} topochemical ion exchange producing closely related but \textit{non-isostructural} derivatives. The existence of Li$_x$BC precursors greatly expands the space of viable products to non-stoichiometric ternary or quaternary derivatives and can aid in the rational design of superconductors.

Remarkably, the accessible stoichiometric AgBC with the simplest considered three-atom unit cell already embodies the long-sought features for conventional superconductivity. Compared to the MgB$_2$, CaC$_6$, and predicted LiB superconductors, AgBC benefits from all three e-ph coupling channels present in honeycomb compounds. According to our aME calculations, the more balanced contributions result in $T_{\rm c}$ values reaching 56~K. Modest hole-doping could further boost $T_{\rm c}$ by turning ellipsoidal BC-$p_{x,y}$ Fermi sheets into hallmark 2D cylinders. CuBC is predicted to be a frustrated metal with low DOS at the Fermi level, but topochemical synthesis from Li$_x$BC may produce even better Cu$_x$BC superconductors. Altogether, the reintercalation and topochemical pathways proposed in our previous~\cite{Kharabadze2023,Tomassetti2024a,Tomassetti2024b} and present studies, respectively, establish layered metal borocarbides as primary candidates to superconduct at near or above liquid nitrogen temperature.

\begin{acknowledgments} 
We thank Igor Mazin for valuable discussions and Christopher Renskers for his help in identifying the CuBC structural model. The authors acknowledge support from the National Science Foundation (NSF) (Award No. DMR-2132589). This work used the Expanse system at the San Diego Supercomputer Center via allocation TG-DMR180071 and the Frontera supercomputer at the Texas Advanced Computing Center via the Leadership Resource Allocation (LRAC) award DMR22004. Expanse is supported by the Extreme Science and Engineering Discovery Environment (XSEDE) program~\cite{XSEDE} through NSF Award No. ACI-1548562, and Frontera is supported by NSF Award No. OAC-1818253~\cite{Frontera}.

\end{acknowledgments}


\bibliography{refs} 

\end{document}